\documentclass{aa}
\usepackage{color}
\usepackage{graphicx}
\usepackage[fleqn]{amsmath}	
\usepackage{amssymb}	
\usepackage{mathtools}
\usepackage{txfonts}
\usepackage{amsmath}
\usepackage{natbib,twoopt}

\usepackage{float}
\usepackage[breaklinks=true]{hyperref}
\hypersetup{
  colorlinks=true,   
  citecolor=blue,    
  urlcolor=blue,     
  linkcolor=blue,     
}
\bibpunct{(}{)}{;}{a}{}{,} 

\newcommand{\lcdm}{$\Lambda$CDM\xspace}

\newcommand{\mbh}{$M_{\rm BH}$\xspace}
\newcommand{\lbol}{$L_{{\rm BOL}}$\xspace}
\newcommand{\ledd}{$\lambda_{{\rm EDD}}$\xspace}

\begin{document} 

   \title{A thin disk and a nearly universal accretion rate in luminous quasars}
   
   \author{
    G. Risaliti\inst{1,2}\thanks{\email{guido.risaliti@unifi.it}}, B. Trefoloni\inst{2,3}, M. Salvati\inst{2}          }

\institute{
$^{1}$ Dipartimento di Fisica e Astronomia, Universit\`{a} di Firenze, via G. Sansone 1, 50019 Sesto Fiorentino, Firenze, Italy\\
$^{2}$ INAF-Osservatorio Astrofisico di Arcetri, Largo E. Fermi 5, 50125, Firenze, Italy\\
$^{3}$ Scuola Normale Superiore, Piazza dei Cavalieri 19, 56126, Pisa, Italy.
}
   \date{\today}

  \abstract{Quasars accretion models predict a broad range of optical and ultraviolet properties that depend primarily on black hole mass and accretion rate. Yet, most optically selected luminous quasars display strikingly similar continuum spectra. We show that this uniformity can be explained by a nearly constant luminosity-to-mass (“Eddington”) ratio, \ledd and by thermal emission from a standard, optically thick, geometrically thin accretion disc. A standard disk with an Eddington ratio \ledd$\sim$0.1 reproduces both the black hole mass–luminosity distribution of Sloan Digital Sky Survey (SDSS) quasars and their principal continuum properties. In this framework, the spectral energy distribution (SED) peaks beyond the observable ultraviolet range for nearly all sources. We show that the few quasars, expected to be cold enough to shift the peak into the observable region, indeed show this behavior. This scenario is further supported by an analysis of the relation between the luminosity of the main broad emission lines and the continuum luminosity (i.e. the `Baldwin effect'). We find that 1) the observed slopes of the line--continuum relations match the expectations from the standard disk model, if we assume that the line emission is a good proxy of the ionizing luminosity; 2) the dispersions of the line-continuum luminosity relations are very small (as small as 0.13 dex), suggesting that the physics of the disk-broad line region system is dominated by only one parameter (the black hole mass) with a nearly constant Eddington ratio. Finally, we notice that our hypothesis of constant \ledd$\sim$0.1 provides a black hole mass estimate (based on the observed luminosity) with a smaller error than the virial estimate. } 

\titlerunning{A new quasar sample for cosmology}
\authorrunning{G. Risaliti et al.}
   \maketitle
%

\section{Introduction}
Quasars are the most luminous persistent sources in the Universe and are powered by accretion onto a supermassive black hole (mass $M_{BH}>10^6~M_\odot$). Despite a huge amount of observational data, there is no consensus on how the accretion power is turned into electromagnetic radiation. On the theoretical side, the “standard” reference model consists of an optically thick, geometrically thin, accretion disk emitting locally as a black body (\citealt{shakura1973, novikov1973}). This model has both significant strengths and weaknesses. Its main successful predictions are (a) a high efficiency in converting accreting mass into energy, which is required to explain the high ratio between the total luminosity emitted by quasars and their relic masses (the so-called “Soltan argument” (\citealt{soltan1982, marconi2004}); and (b) a broad agreement with the observed Spectral Energy Distribution (SED) of quasars, peaking in the UV. The main theoretical limitations of the model lie in its simplified treatment of viscosity and energy transport, and on its (lack of) self-consistency at either low or high accretion rates. In particular, using the “Eddington ratio” \ledd~ as a measure of the accretion rate, values \ledd$<0.01$ imply an inefficient, advection-dominated flow, while at \ledd$>0.2-0.3$ the radiation-dominated inner part of the flow can no longer maintain a geometrically thin shape and a “slim” disk \citep{abramowicz1988} and/or strong radiation-driven outflows can easily develop \citep{laor2014}. Observationally, shortcomings of the standard model are the observed variability time scales \citep{lawrence2018}, which is too fast to be explained by viscous processes; the size inferred from variability and micro-lensing measurements, which is a factor of 3-4 larger than expected (\citealt{morgan2010, fausnaugh2016}); and the direction of the observed polarization \citep{stockman1979, antonucci1988}. A disk made of optically thick “blobs” interconnected by a tenuous, semi relativistic inter-blob medium could cure most of these. A more serious problem is that the SED does not fully match the model predictions. The latter point has been the subject of several recent studies: the standard model has been applied to the quasar spectra of the Sloan Digital Sky Survey \citep{york2000}, whose latest catalogue (the DR16, \citealt{wu2022}) contains about 750,000 quasars, to compare the expected spectral shape with the observations. Specifically, a broad expectation of any geometrically thin, optically thick disk, locally emitting as a black body, is that the monochromatic luminosity $L_\nu$ scales with the frequency as $L_\nu\sim\nu^{1/3}$. This holds up to a value $\nu_{MAX}$ determined by the maximum temperature of the disk, $T_{MAX}$. Such a temperature, in turn, depends on the luminosity and on the black hole mass $M_{BH}$; while the former is directly observed, the latter is usually estimated through the so-called “virial relations” (see Methods for details). According to this procedure, the SED of most quasars is expected to peak (or reach its maximum) at wavelengths shorter than the Lyman limit, therefore making the direct measurement of $T_{MAX}$ impossible. However, a subset of them should have $\nu_{MAX}$ within the spectral range accessible to the SDSS, but the observations do not confirm these expectations: all the stacked spectra have a nearly indistinguishable continuum shape, regardless of the predicted peak temperature (e.g. \citealt{mitchell2023, trefoloni2024}). This observational result, combined with new UV photometric studies (e.g. \citealt{cai2023}), led some authors to abandon the standard disk hypothesis altogether. A nearly universal spectrum, similar to a disk spectrum but always peaking around the Lyman limit, could be produced by a combination of ionization and transport processes. Alternatively, the emission around the peak could be due to Comptonization of a primary soft X-ray source \citep{kubota2018}. 

The starting point of this paper is the consideration 
that all the above results are strongly dependent on the value of the black hole mass $M_{BH}$ and must be reconsidered if the uncertainties on such measurements are properly taken into account. 
In Section~2 we analyze the mass and Eddington ratio \ledd~ distributions of the SDSS quasars and we argue that they are consistent with a very narrow distribution of \ledd.
In Section~3 we test the hypothesis of a nearly constant Eddington ratio on the stacked spectra of SDSS quasars. In Section~4 we analyze the Baldwin effect of several broad emission lines and show that it can also be easily explained through a standard disk model with a nearly constant Eddington ratio. We discuss and summarize our results in Section~5.

\section{The \mbh~ and \ledd~ distributions in SDSS quasars}

Quasar mass estimates are mostly based on the “virial” method, i.e. the assumption that the width of the broad lines is due to the Keplerian motion of the emitting clouds around the black hole, and that the ionization state of such clouds is about the same for all quasars. With these assumptions, we expect a relation between the black hole mass $M_{BH}$, the line width $W$ and the source luminosity $L$: 
$M_{BH} = K\times W^{2} L^\alpha$,
where $\alpha$ and $K$ are parameters that can be fitted to the data ($\alpha$=0.5, if the ionization state is exactly the same in all quasars). The calibration has been obtained from sources with independent black hole mass measurements from reverberation mapping \citep{vestergaard2006, park2012, dallabonta2020}: in all these works, the estimated uncertainty in the virial masses is of the order of at least 0.4-0.5 dex (see \citealt{shen2013} for a detailed discussion on this point). This often overlooked high uncertainty has several immediate implications. 
\begin{figure}[h!]
\centering
\includegraphics[width=\linewidth,clip]{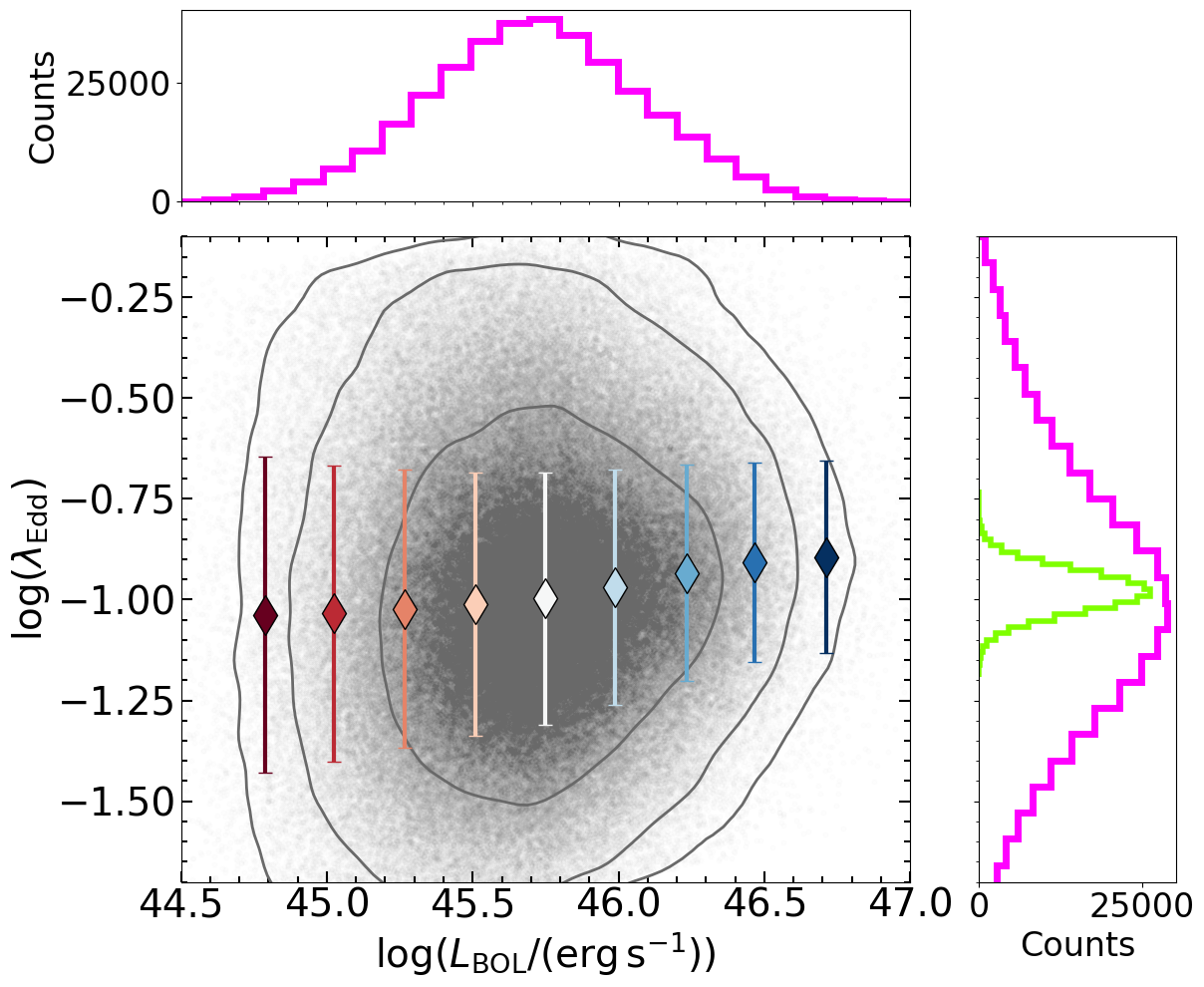}
\caption{Eddington ratio versus bolometric luminosity for a sample of SDSS quasars. The luminosities are obtained from the monochromatic values of \citet{wu2022} assuming a standard disk model, while the black hole masses are estimated with the virial method \citep{vestergaard2006}. The diamonds show the mean \ledd~ values in small luminosity bins, with error bars showing their dispersions. The total distributions are shown as magenta histograms in the side panels. The green histogram shows the “intrinsic” distribution of \ledd needed to reproduce the small increase of \ledd with luminosity. The width of this distribution is $\sigma$=0.05 dex. Contours enclose 68, 95 and 99\% of the underlying population. }
\label{fig:Ledd-Lbol}
\end{figure}

In Fig.~\ref{fig:Ledd-Lbol}, we plot the distribution of \ledd~ and the bolometric luminosity \lbol for $\sim$100,000 quasars in small luminosity bins between $z=0.8$ and $z=2$. Data are taken from the analysis of \citet{wu2022}, with the following prescriptions:\\
1) We choose quasars in the redshift range $z=0.8-2.0$, where the Mg~II~$\lambda~2800$~\AA~ line is present, and we select sources with a line width FWHM$>1,000$~km/s.\\
2) We adopt the estimate of the virial mass of \citet{wu2022}, which uses the \citet{shen2011catalog} calibration.\\
3) We estimate the total disk luminosity $L_{\rm BOL}$ from the monochromatic one, $L_\nu$, at 3,000~\AA~ assuming a standard optically thick and geometrically thin accretion disk. 
Within this model, an excellent approximation of the total disk luminosity is  $\nu_{\rm MAX}L_{\nu,{\rm MAX}}$, where $\nu_{\rm MAX}$ is the peak frequency of the spectrum. Assuming a spectral shape $L_\nu\sim\nu^{1/3}$ we have $L_{\rm BOL}\sim (\nu_{\rm MAX}~L_{\nu,{\rm MAX}}\propto \nu_{\rm MAX}^{4/3})$.

A visual inspection of Figure~\ref{fig:Ledd-Lbol}, as well as subsequent fits of the distributions, show that the total distribution of \ledd~ has a dispersion of 0.45 dex, while the individual distributions at fixed luminosity have dispersions of the order of 0.35~dex, with a slight increase of the average value of \ledd~with the luminosity. Even considering (1) only the total distribution, (2) the smallest estimated uncertainty in $M_{BH}$, and (3) a negligible error in the estimate of the bolometric quasar luminosity $L_{BOL}$ and recalling that \ledd$\sim~L_{\rm BOL}/M_{\rm BH}$, we conclude that there is no room for an intrinsic dispersion of \ledd~greater than 0.1~dex. In particular, an intrinsic dispersion of 0.05~dex is enough to include the small increase with luminosity of the average values, and can easily reproduce the estimated total dispersion when a black hole mass uncertainty of 0.4~dex is added.\\
The constraint is even tighter if the distributions at fixed luminosity are considered: in this case, we need to assume an almost fixed value of \ledd and an uncertainty on $M_{BH}$ even smaller than that estimated in the calibration works \citep{vestergaard2006, park2012, dallabonta2020}.\\

Based on the observation that the intrinsic \ledd distribution must be tight, there are at least three physically and observationally far-reaching consequences. \\
\begin{itemize}
\item If \ledd~ is nearly constant (with a dispersion of the order of, or lower than, 0.1~dex) and with an average value $\langle\log(\lambda_{EDD})\rangle\sim -1$ so as to match the average of the \ledd~distribution estimated using the virial masses in the SDSS sample, the number of very low-\ledd or very high \ledd~(up to possibly super-Eddington) is negligible. \\
\item Within this framework, due to the small inferred scatter in \ledd, the luminosity itself is a better indicator of the black hole mass than the “virial” estimates \\
\item The power source of the quasars in the SDSS catalog is expected to be dependent only on one main physical parameter, i.e. the black hole mass, spanning more than three orders of magnitude, with the other parameters (\ledd, and possibly the black hole spin) only playing a minor role in determining their physical state and their emission properties. 
\end{itemize}
In the following Sections, we address these three points and quantitatively test the scenario of a nearly constant Eddington ratio in two different ways: the prediction of the spectral shape of SDSS quasars and the line-to-continuum luminosity relation (the so-called “Baldwin effect”, \citealt{baldwin1977}).

As a working hypothesis, we will assume \ledd=0.1. This is the average value of SDSS quasars adopting the virial mass estimates. We will discuss the implications of this choice in the next Sections and in the Conclusions.

\begin{figure*}[h!]
\centering
\includegraphics[width=\linewidth,clip]{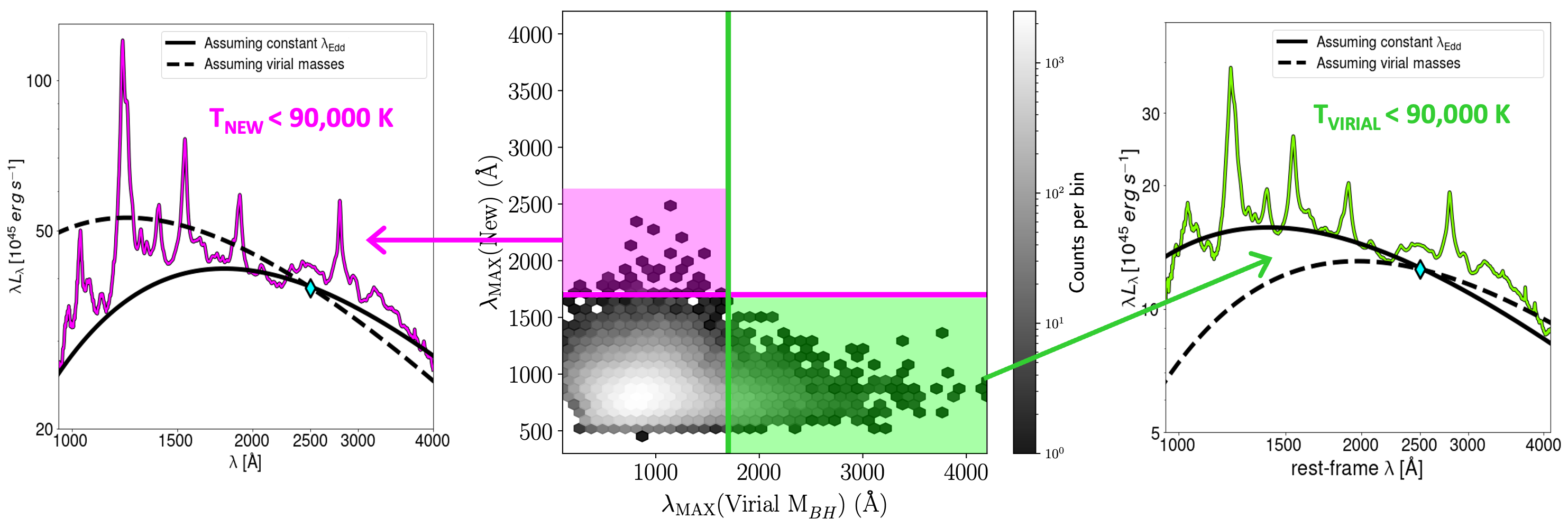}
\caption{\textit{Central panel:} expected peak wavelength of the quasar spectrum based on virial masses (x-axis) and on our assumption of constant Eddington ratio (y-axis). The pink and green lines show the maximum value $\lambda\sim1,600$~\AA~ (corresponding to a maximum disk temperature $T_{MAX}=90,000~K$) required to see a clear spectral change with respect to the disk power law in the observed UV spectrum, i.e. at wavelengths shorter than the Lyman limit. While the virial masses predict a significant fraction of “cold” objects that should show the emission peak in the observed spectral range (and it is not seen), we predict that only very few quasars are cold enough to show the peak. \textit{Right panel:} stacked spectra of SDSS quasars with expected peak temperatures $T_{MAX}> 90,000~K$. The values of $T_{MAX}$ are obtained from a standard disk model with a black hole mass $M_{BH}$ estimated with the virial method. The dashed line shows the expected continuum spectrum based on these assumptions, superimposed (not fitted) on the data and rescaled to the average continuum intensity at 2,500~\AA, as estimated in the SDSS DR16 quasar catalogue. The solid line shows the expected average spectrum assuming a constant Eddington ratio \ledd=0.1. \textit{Left panel:} the same, with $T_{MAX}$ estimated from a standard disk model with \ledd=0.1. The mismatch between the observed spectra and the predictions based on the virial masses is obvious. Conversely, the average spectra with \ledd=0.1 are in good agreement with the observed data.}
\label{fig:moneyplot}
\end{figure*}

\section{A solution to the "Universal SED" puzzle}
In Fig.~\ref{fig:moneyplot}, we show the predicted wavelength of the emission peak for SDSS quasars, assuming a standard disk and a black hole mass as derived with our method (i.e. assuming a constant \ledd=0.1) and, alternatively, with the virial method. Our tests on SDSS spectra showed that, to produce a detectable slope change in the available spectral range, the emission peak must be at a rest-frame wavelength longer than $\sim$1,700~\AA. Fig.~\ref{fig:moneyplot} clearly shows that the quasars predicted to be cold enough to exhibit a slope change are two disjoint sets depending on which method is used to estimate $M_{BH}$. 

In order to observationally test the two scenarios, we stacked the “cold” quasars according to the two predictions and compared them with the corresponding models. The procedure is analogous to that described in \citet{lusso2015first}. Here we briefly outline the main steps performed to build the composite SDSS spectrum in each temperature bin:
\begin{itemize}
\item Each spectrum was corrected for Galactic absorption assuming the value for the color excess E(B-V) available in the catalog from \citet{wu2022} according to the extinction maps of \citep{schlafly2011} and the extinction curve derived in \citep{fitzpatrick1999}. Then, the de-reddened spectra were shifted to the rest frame.\\
\item All the spectra in the same sample were resampled, by means of linear interpolation, onto a logarithmically equally spaced wavelength grid. Successively all the spectra were scaled by their mean flux in a small wavelength interval ($\sim$40~\AA) close to 2500~\AA, in order not to bias the stack towards the most luminous objects in each subsample.\\
\item Defective spectra (e.g. those with less than 10\% of pixels flagged as `good’), or spectra where the rest frame 2500~\AA~ emission was not covered, were discarded. The fraction of objects rejected during this step was of the order of 2\%.
\end{itemize}
The final composite spectrum was obtained by taking the median value of the flux distribution in each spectral channel. The uncertainty on the median value was evaluated as the semi-interpercentile range between the 1st and the 99th percentiles in each spectral channel divided by the square root of the number of sources contributing to that channel. Since during the stacking process the spectra had been normalized, we bring the composite back to physical units by scaling it to the average luminosity at 2,500~\AA~ of the subsample.

The results clearly confirm the constant \ledd scenario: the “cold” quasars according to the virial estimates of $M_{BH}$ are not observed to be cold, while the ones predicted to be cold based on the constant \ledd hypothesis indeed show the expected spectral peak. We also notice that the truly “cold” quasars are only a few tens out of more than 100,000 sources. It is therefore not surprising that they had not been discovered before: almost all quasars in the SDSS sample have a broad peak reaching its maximum at wavelengths shorter than the Lyman limit. Their spectral properties are remarkably similar in the observable spectral range (partly owing to the SDSS selection pipeline), and the few exceptions are not captured by the virial mass method. In our scenario, the few cold-enough quasars are simply the most luminous ones. We notice that a recent analysis of the most luminous known quasar (J2157-3602 at $z$=4.7; \citealt{wolf2024accretion}) also shows a very "cold" UV spectrum and a sub-Eddington accretion rate, in agreement with our expectations.

\begin{figure}[h!]
\centering
\includegraphics[width=\linewidth,clip]{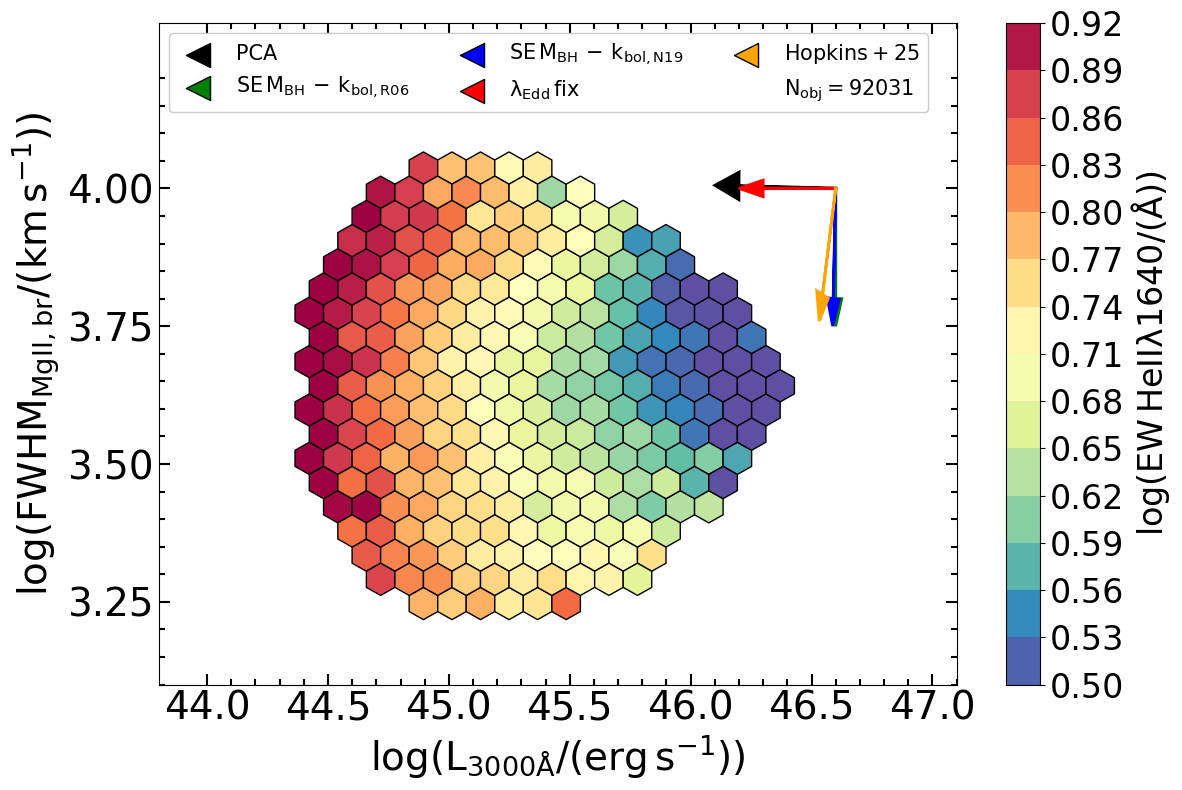}
\caption{Test of disk models based on three observables in quasar spectra: the continuum luminosity at 3,000~\AA, the FWHM of the Mg~II~$\lambda$~2800~\AA\, line and the equivalent width of the high ionization He~II~$\lambda$~1640~\AA\, line. Assuming that the equivalent width of the He~II is proportional to the ionizing luminosity it is possible to predict its dependence on the black hole mass and Eddington ratio in several physical scenarios: a standard disk assuming virial black hole masses and a constant (e.g. R06; green arrow) or luminosity-dependent bolometric correction (blue arrow); the accretion disk model of \citet{hopkins2025} (orange arrow); a standard disk with constant \ledd (red arrow). The latter scenario is the only one in agreement with the observational results obtained via the PCA and the linear regression.}
\label{fig:L_W_He2}
\end{figure}

An analogous way to prove the ability opf the proposed approach at predicting the disk temperature in our quasar sample is shown in Fig.~\ref{fig:L_W_He2}. There, we present the equivalent width of the broad He~II~$\lambda$~1,640~\AA\, line in the $\rm \log(L_{3000\AA})$--$\rm \log(FWHM_{MgII})$ parameter space. We chose the EW of the UV He~II line, because being a `photon-counter' it represents an excellent proxy of the total ionizing flux (see e.g. \citealt{ferland2020state, temple2023testing}). As increasingly colder accretion disks produce fewer ionizing photons, we expect that the evolution of the He~II EW with the black hole mass and luminosity informs us about the underlying SED temperature. Abiding by our data-driven approach, we adopted an observational parameter space, using the FWHM of the Mg~II~$\lambda$~2800~\AA\, line and the 3,000~\AA\, monochromatic luminosity as axes. This parametrization allows us to explore the expected direction of increasing temperature according to different prescriptions. A direct test of our framework on a subsample of SDSS quasars, by directly employing accretion disk modelling, in this same parameter space will be subject of a forthcoming work (Trefoloni et al., in prep.).

We estimated the direction of strongest correlation via a partial correlation analysis (PCA; see e.g. Sec.~5.2 in \citealt{trefoloni2025missing} for more details). In brief, the ratio of the Spearman correlation coefficients between the He~II EW and the quantities along the y- and x-axes defines the tangent of the PCA angle, whose direction is shown as a black arrow. To compare the observed direction with those predicted by different models, in Fig.~\ref{fig:L_W_He2} we show the directions of increasing temperature according to different assumptions on the relation between \mbh and \lbol. In particular, we show the line of increasing temperature for a standard disk (Eq. \ref{eq:tmax}) assuming that the mass is given by single-epoch virial calibrations and a fixed bolometric correction (\citealt{richards2006spectral}; R06) as a green arrow. Virtually the same direction is found assuming a luminosity dependent bolometric correction (\citealt{netzer2019bolometric}; N19, blue arrow).

By re-arranging  Eq. \ref{eq:tmax} and~2, and by writing the $EW$ as $L_{\text BOL}/L_\nu$ (see disxcussion below), one finds $T_{\rm MAX}\sim \lambda_{\rm EDD}^{1/2} \, L_{\rm BOL}^{-1/4} \sim {\text FWHM}^{-1}$; and $EW \sim L_\nu^{-1/4} \, \lambda_{\text EDD}^{1/2} \sim {\text FWHM}^{-4/3}$. If the Eddington ratio has indeed a narrow distribution, implying the irrelevance of the virial approach, the temperature and the He~II EW vary only along the luminosity axis. It is clear, by comparing the arrows in the top right corner that the constant \ledd model (red arrow) is by far the preferred one. For completeness, there we also show the expected dependence of $T_{\rm MAX}$ on the black hole mass and luminosity, assuming single.epoch calibrations for a magnetically-dominated accretion disk as discussed in \citet{hopkins2025}. Finally, we remark that directly computing the coefficients ($a$, $b$) of a multiple linear regression in the form $\rm{EW \, HeII}= a \log(L_{3000\AA}/(erg \, s^{-1})) + b \log(FWHM_{MgII}/(km \, s^{-1})) + c$ using \textsc{emcee} (see Sec.~\ref{sec:baldwin} for more details) instead of adopting the PCA, provided analogous results.

\section{A reanalysis of the Baldwin effect}
\label{sec:baldwin}
A second powerful way to obtain information about the disk structure and emission is the comparison between the intensity of the broad emission lines and the continuum. The broad emission lines in quasars are emitted by photo-ionized gas rotating around and/or outflowing from the central black hole, with velocities in the range 1,000-10,000 km/s. Their detailed properties depend on the ionizing continuum shape and intensity, and on its complex interaction with the emitting gas, as modeled by several photo-ionization codes. 
The spectroscopic measurements of the line and continuum parameters are derived from the analysis by \citet{wu2022}. We analyzed the line to continuum relation for six broad emission lines: 
He~II~$\lambda~4687$~\AA, Mg~I~$\lambda~2800$~\AA, C~III]~$\lambda~1909$~\AA, He~II~$\lambda~1640$~\AA, C~IV~$\lambda~1549$~\AA. 
\begin{table*}[h!]
\centering
\begin{tabular}{lccccc}
\hline
Line & $N$ & $\lambda$ (\AA) & $\langle \gamma \rangle_{\mathrm{obs}}$ & $\gamma_{\mathrm{model}}$ & $\langle \delta_{\mathrm{INTR}} \rangle$ \\
\hline
Mg\,II $\lambda$ 2800\,\AA & 128815 & 3000 & 0.74$\pm$0.01 & 0.73 & 0.15 \\
C\,III] $\lambda$ 1909\,\AA & 56789 & 1700 & 0.70$\pm$0.01 & 0.72 & 0.13 \\
C\,IV $\lambda$ 1549\,\AA & 98164 & 1700 & 0.64$\pm$0.01 & 0.66 & 0.21 \\
He\,II $\lambda$ 1640\,\AA & 6553 & 1700 & 0.69$\pm$0.02 & 0.68 & 0.19 \\
Si\,IV + O\,IV] $\lambda$ 1398\,\AA & 63068 & 1700 & 0.64$\pm$0.01 & 0.65 & 0.20 \\
He\,II $\lambda$ 4687\,\AA & 535 & 5100 & 0.74$\pm$0.03 & 0.77 & 0.26 \\
\hline
\end{tabular}
\caption{Results of the line--continuum fits for all the broad emission lines measured in at least 500 SDSS quasars. For each line we report the number of available objects, the wavelength of the continuum luminosity used in the fit, the average slope of the line--continuum relation, the expected slope based on a thermal disk model (assuming that the line is a proxy for the total luminosity at wavelengths shorter than that of the line), and the intrinsic dispersion of the relation.}
\label{tab:eddington}
\end{table*}
These are all the brightest emission lines in the typical spectrum of a luminous quasar. We have neglected the hydrogen lines Ly~$\alpha$, H$\alpha$ and H$\beta$, whose luminosities are heavily dependent on the details of energy transport within emitting clouds, and are therefore less directly related to the ionizing luminosity. 
The continuum luminosity has been estimated at 1,700~\AA~ and 3,000~\AA, depending on the emission line, as shown in Table~1. We chose the closest available continuum wavelength to each line.  
For each subsample, we filtered SDSS quasars only requiring a redshift range fully containing both the relevant line and continuum, and a continuum flux measurement with an error lower than 10\%. Furthermore, we excluded radio-loud and broad-absorption-line (BAL) quasars.
For each line, we performed a linear fit in a log-log plane in several narrow redshift bins, with $\Delta[\log(z)] < 0.03$. This allows us to investigate a possible evolution of the slope of the relation, without assuming any cosmological model adopted to derive luminosities from fluxes, since the differences among distances in each bin are much smaller than the dispersion of the relation (see, e.g. \citealt{risaliti2019cosmological} and references therein. The importance of this approach is discussed in Appendix~\ref{app:supp_figures}, where we show that the possible change with redshift of the normalization of the line to continuum relation can lead to incorrect estimates of the slope. To obtain the best fit parameters and their errors, we maximized a customized likelihood including an intrinsic dispersion of the data with respect to the best fit relation, and the contributions of both the line and continuum measurement errors. The minimization was performed with the Bayesian Monte-Carlo code \textsc{emcee} (\citealt{foreman2013emcee}). We also checked that in no cases the residuals show a significant deviation from a purely linear fit. Examples of the data and best fits are shown in Figure~4 and in Appendix~\ref{app:supp_figures}. 
\begin{figure}[h!]
\centering
\includegraphics[width=\linewidth,clip]{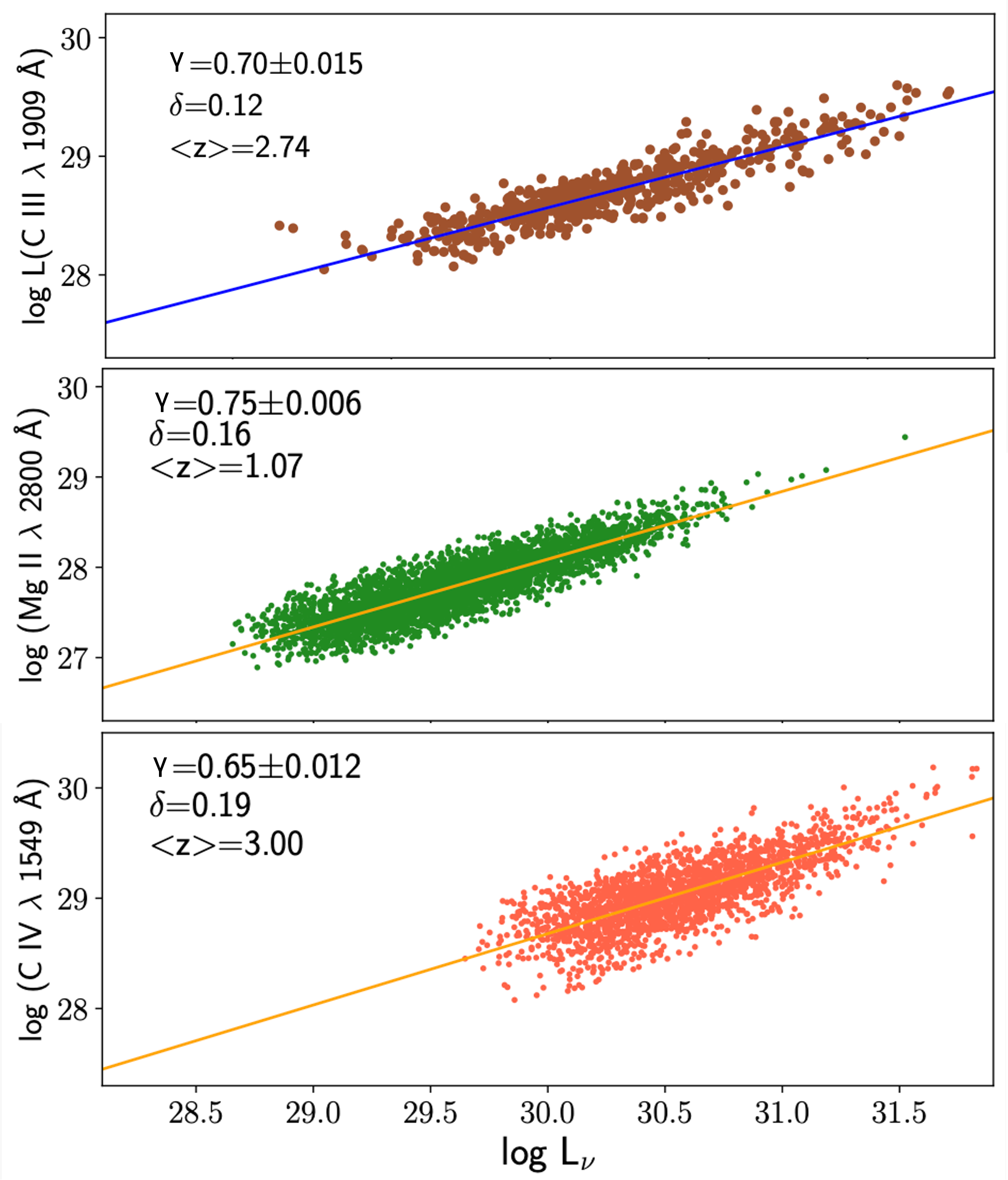}
\caption{Examples of line-continuum relations for three broad emission lines (Mg~II~$\lambda 2800$~\AA\, in the upper panel, C~III]~$\lambda~1909$~\AA\, in the middle panel, C~IV~$\lambda$~1549~\AA\, in the lower panel) in three small redshift intervals ($\Delta[(\log(z)]\sim0.03$. The values in the labels refer to the best-fit slope, $\alpha$, and the intrinsic dispersion $\delta$ with respect to the linear log-log relation.}
\label{fig:examples}
\end{figure}

\begin{figure*}[h!]
\centering
\includegraphics[width=0.9\linewidth,clip]{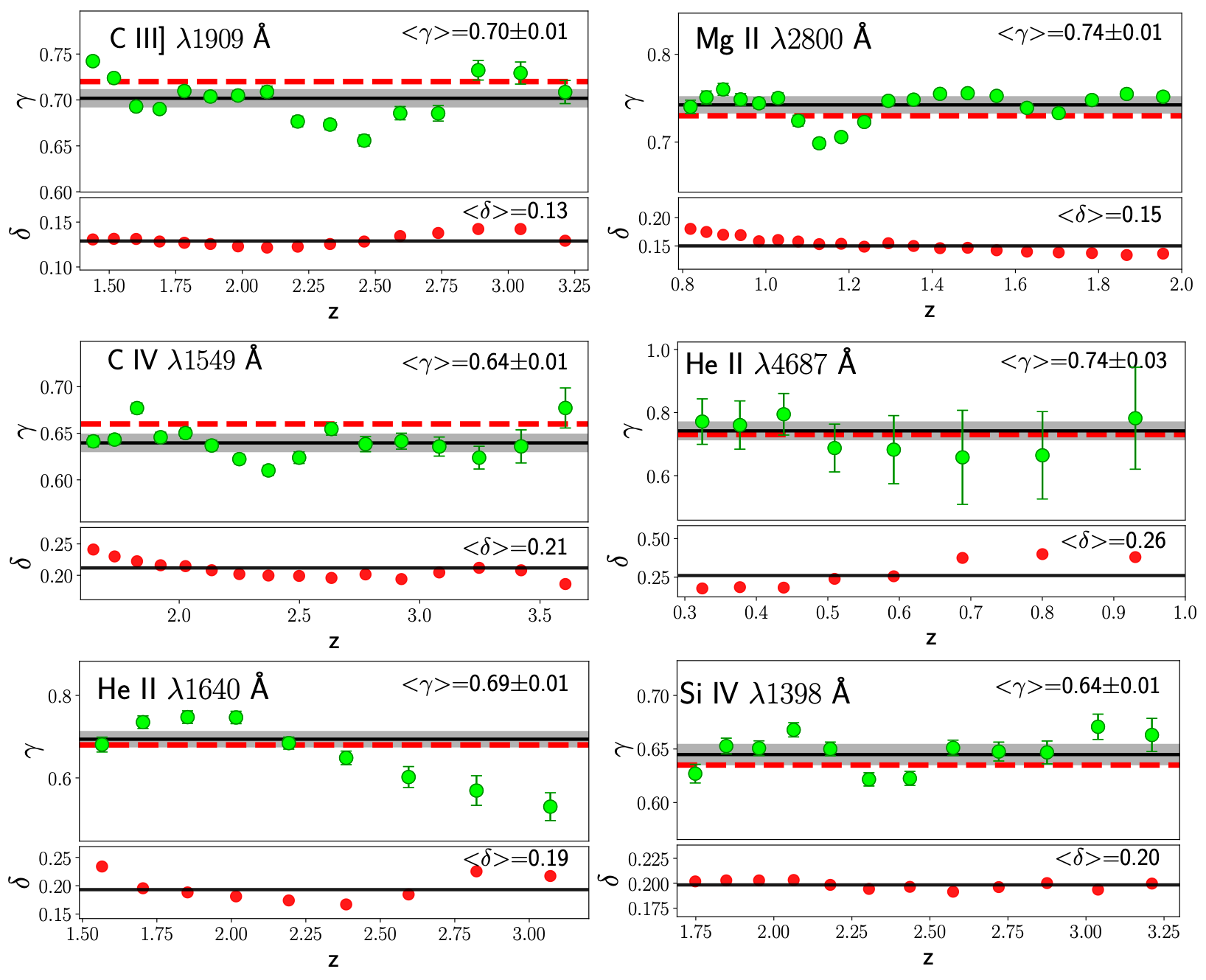}
\caption{Best fit slopes as a function of redshift for all the analyzed emission lines. The black lines and the grey region show the average slope and its uncertainty for each line. The red dashed lines show the expectation based on a standard disk model (see text for details). The lower part shows the intrinsic dispersion in each redshift bin.}
\label{fig:BaldwinTOT}
\end{figure*}

\begin{figure}[h!]
\centering
\includegraphics[width=\linewidth,clip]{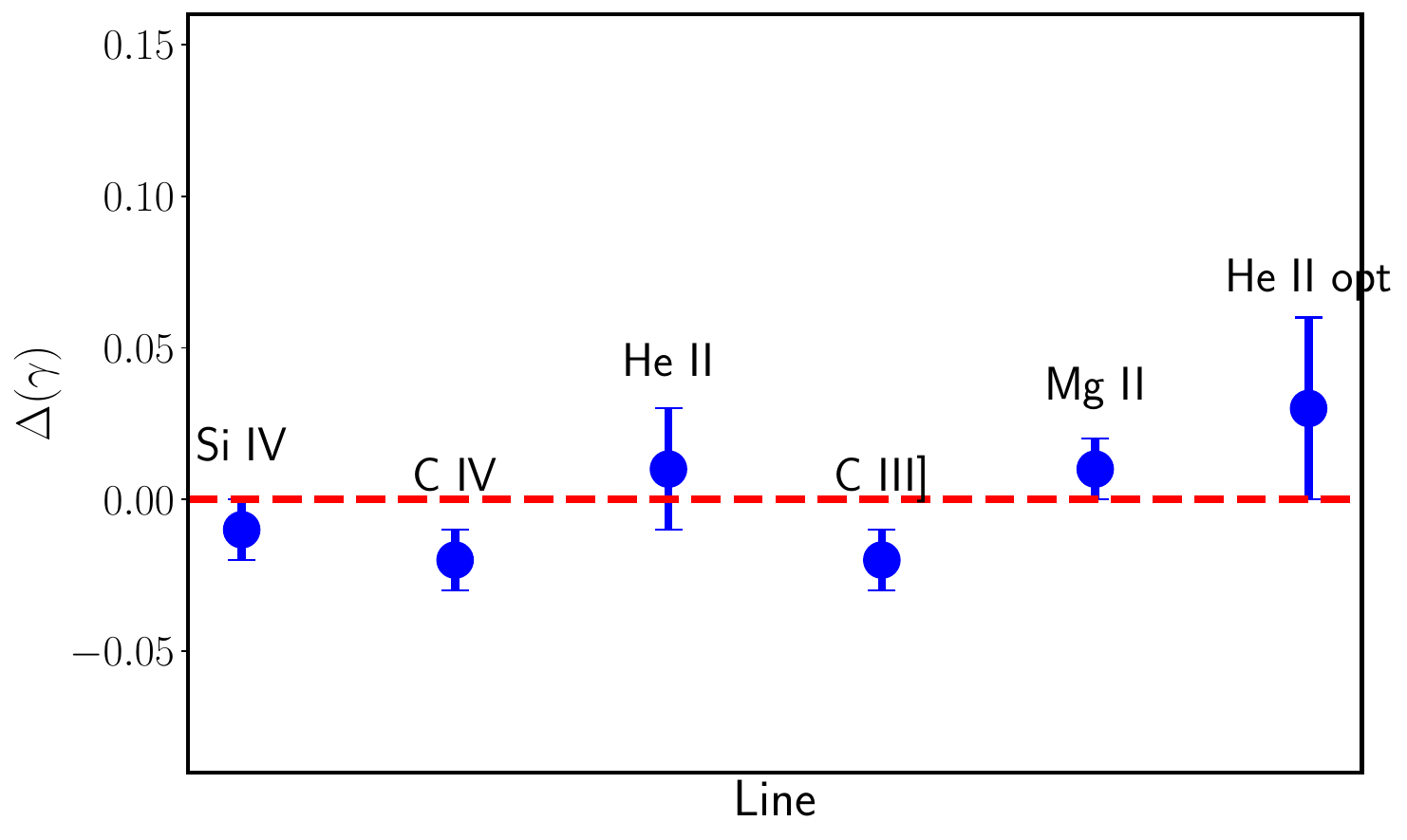}
\caption{Difference between the average slope and the expectation from the standard disk model for the main broad emission lines.}
\label{fig:Summary}
\end{figure}

The main results of the analysis are shown in Table~\ref{tab:eddington} and Figure~\ref{fig:Summary} and are described below:
\begin{itemize}
\item The line-to-continuum slope has exponents in the range 0.65-0.75. A value of 0.75 is expected within the standard disk model in the relation between the total (\lbol) and monochromatic ($L_{\nu}$) disk luminosity. This is easily demonstrated by assuming the standard disk spectral shape $L_\nu = k \nu^{1/3}$ and that $L_{\rm BOL} = \nu_{\text MAX} L_{\nu, \text{MAX}} = k \nu_{\text MAX}^{4/3}$. The peak frequency is related to the disk parameters as:
\begin{equation}
\nu_{\text{MAX}}\sim T_{\text{MAX}}\sim \left[\frac{M_{\text{BH}}L_{\text{ BOL}}}{R_{\text{IN}}^3}\right]^{\frac{1}{4}}
\label{eq:tmax}
\end{equation}
where $R_{\rm IN}=\eta~GM_{\rm BH}/c^2$ is the inner radius of the disk, 
with the parameter $\eta$ including the possible relativistic effects. Under these assumptions, it is immediate to obtain a relation: 
\begin{equation}
L_{\rm BOL} \sim L_\nu^{3/4} \lambda_{\rm EDD}^{1/2} \eta^{-3/4}
\label{eq:Lbol_Lline}
\end{equation}
If the luminosity of the emission lines is roughly proportional to the total luminosity, the same exponent is expected in the line-monochromatic luminosity relation. 
The agreement between the disk prediction and the observational results increases if a numerical calculation is adopted: we simulated the standard disk emission assuming that it extends down to an inner radius $R_{IN}=6GM_{BH}/c^2$ and neglecting general relativistic effects (we checked with a KERRBB model, \citealt{li2005multitemperature}, that their inclusion does not significantly alter our results). For each line we considered a luminosity distribution as measured in each of the observed subsamples, with a constant Eddington ratio log(\ledd)=-1 and computed the continuum monochromatic luminosity indicated in Table 1 and the total ionizing luminosity for each line. The latter defined as the entire continuum emission at wavelengths shorter than that of the line. We then performed a linear fit of these two quantities, obtaining the values reported in Table~\ref{tab:eddington}. The (small) differences with the zero-th order estimate of 0.75 are due to the continuum slope not being exactly 1/3 at the continuum frequency (especially for relatively cold sources) and to the ionizing luminosity not being the same as the total disk luminosity. The encouraging agreement between these estimates and the measured slope values is shown in Figure~\ref{fig:BaldwinTOT}. \\
\item The dispersion of the relation is very small: 0.13 dex for C III], 0.14 for Mg II (but even smaller at z $>$ 1). 

There are at least three expected contributions to this dispersion, regardless of the physical model used to interpret the data:\\
1) Inclination of the system. Even without specific assumptions on the structure of the emitting source, we can expect, on very general grounds, that the line emission is optically thin, therefore emitted isotropically, while the continuum emission comes from an optically thick medium. This necessarily implies an orientation-dependence of the continuum luminosity, unless the source is completely spherically symmetric. Considering a random distribution of the inclination of the objects in the sky and a selection effect in favor of face-on objects (due to the flux-limited selection of the SDSS quasars), it has been shown that the expected dispersion due to orientation is of the order of 0.25 dex, far exceeding the observed dispersion in the $L_{\rm LINE}$-$L_\nu$ relation \citep{signorini2024}. The presence of an obscuring torus, coaxial with the accretion disk, would significantly reduce the observed dispersion.\\ 
2) Relation between the line intensity and the ionizing luminosity. Many studies employing advanced photoionization codes showed that the luminosity of a given emission line does not scale linearly with the ionizing luminosity (see e.g. \citealt{ferland2020state} and references therein). The ionization structure of the emission clouds also plays a role: for example, an increase in the total UV luminosity may over-ionize an emitting cloud and reduce the luminosity of low-ionization ions. Again, these non-linear effects are expected to be present, regardless of the details of the emitting source. We expect these effects to introduce a dispersion with respect to a simple linear, or power law dependence on the ionizing luminosity. Even if it is not easy to exactly quantify this effect, it should contribute to the observed dispersion.\\
3) Variability. All the continuum variability modes with periods shorter than the light crossing time to the BLR introduce a further contribution to the observed dispersion. In principle, a reverberation mapping campaign could reduce this effect by providing the optimal time lag between the lines and continuum emission. This has been undertaken for a sizeable quasar sample \citep{shen2024} and will be the subject of further studies.

In addition to these three model-independent effects, a further contribution to the dispersion is expected from the spread in the $\eta$ parameter, which accounts for general relativistic (GR) effects within the $\alpha$-disk scenario (Equation 1):\\
1) The $\eta$ value should be within the interval $\eta$=1 (for a maximally rotating black hole with a co-rotating disk) and $\eta=9$ (for a maximum spin and a counter-rotating disk), with $\eta$=6 corresponding to a non-rotating (Schwarzschild) black hole. However, in a recent study by \citet{hagen2023} a ray-tracing code was used to show that most of the temperature increase due to the decrease of $\eta$ with increasing black hole spin is canceled out by the increase of the relativistic redshift at high spin. Therefore, the effective value of $\eta$ is never lower than $\sim$4. Assuming an average value $\eta$=6 and a spread $\sigma(\eta)=0.1$, the expected contribution to the observed dispersion in Eq.~\ref{eq:Lbol_Lline} is $\sim0.05$~dex.\\

Considering all the above contributions to the dispersion, we should already match, or even exceed, the observed values. However, a further important contribution should come from the width of the \ledd distribution. In principle, if the values in the catalog from \citet{wu2022} (based on the virial black hole mass estimates), are adopted, the spread $\sigma(\lambda_{\rm EDD})$ should be of the order of 0.45, with a contribution to the dispersion in Eq.~\ref{eq:Lbol_Lline} of $\sim0.22-0.23$ dex, much higher than the total observed intrinsic dispersion. 

The main direct conclusion of this analysis is that there is little or no room for a significant width of the distribution of the Eddington ratio \ledd. 
More in general, when such a tight relation is found, the physical system under analysis depends only on one dominant parameter (in this case, the black hole mass). There is no room for a second independent parameter with a range broad enough to introduce a significant scatter in the relation.
\end{itemize}

\section{New black hole mass estimates}
Beyond correctly predicting the shape of quasar spectra and the observed line-to-continuum properties, our hypothesis of a nearly constant Eddington ratio has two far-reaching consequences: 1) it implies that there are no (or a negligible fraction of) super-Eddington sources among SDSS, and 2) the observed luminosity provides a better mass estimate than the virial method. Both statements will be fully discussed in forthcoming publications. Here we just make a few general remarks.

Regarding the Eddington ratio distribution, the identification of a narrow distribution for optically selected quasars does not rule out the case that a broader distribution exists in nature, possibly including super-Eddington sources. Here we only propose that such objects do not show up as blue, unobscured quasars abiding by the selection criteria of the SDSS. For this reason, there is no basis for an extension of our method to quasars with different spectral properties and/or in a different luminosity range, such as the local, low luminosity Seyfert galaxies, or heavily obscured quasars. The same applies to the abundant population of high-redshift, low luminosity AGN discovered by the James Webb Space Telescope (see e.g. \citealt{maiolino2025jwst} and references therein), of which the so-called Little Red Dots represent a noticeable subsample (e.g. \citealt{matthee2024little}). In this context, the extent to which these results can be applied to other samples of optically-selected quasars (e.g. from different surveys) is still under investigation. For instance, recently \citet{trefoloni2025accretion} analyzed a collection of eight high redshift ($z\gtrsim 5$) optically-selected quasars by testing accretion disk models on their SEDs. Albeit with the caveat of a small sample, they found an average $\langle \log(\lambda_{\rm EDD})\rangle \simeq -0.9$ and no evidence for super-Eddington accretion.

\begin{figure}[h!]
\centering
\includegraphics[width=\linewidth,clip]{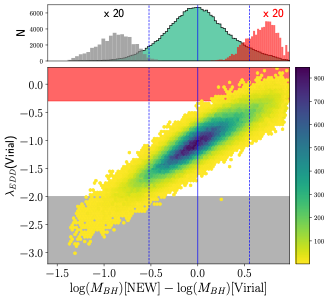}
\caption{Eddington ratio based on virial masses (from \citealt{wu2022}) versus the difference between our new mass estimates and the virial masses. The colorbar shows the number of objects in each bin of the histogram. The upper plot shows the distribution of this difference. The vertical blue lines show the average difference and the 5\% and 95\% percentiles. This shows that for 90\% of the objects the new mass estimates are consistent with the virial values within the 0.5~dex uncertainty of the virial estimations \citep{shen2013}. The grey and red histograms (magnified by a factor of 20) show the distributions of the mass differences for the quasars with \ledd$<0.01$ and \ledd$>0.5$, respectively. For most of these objects the mass difference is higher than 0.5~dex, indicating that the low and high \ledd values among SDSS quasars may be due to an overestimate, or an underestimate, respectively, of \mbh.}
\label{fig:Ledd-MBH}
\end{figure}

Regarding the mass estimates, we note that our values and those based on the virial method are fully consistent for most objects, once the uncertainties in the virial masses are properly considered. Furthermore, as we show in Fig.~\ref{fig:Ledd-MBH}, our narrow distribution of Eddington ratios has the same peak value (by construction) as the one based on virial masses. 
We conclude that the main improvement introduced by our method is “only” a more precise estimate of the black hole masses. The only case where we have significantly different predictions is in the tails of the virial mass distribution, as demonstrated by the average spectra in Fig.~\ref{fig:moneyplot}. 

Finally, it is worth remarking the consequences of our assumption for the physical interpretation of the observational data. On the one hand, we have shown that our hypothesis of a nearly constant Eddington ratio is a convenient way to solve several apparent inconsistencies between the models and the observational results, despite being a major simplification (it removes one of the two main degrees of freedom of the model). On the other hand, this simplification is interpreted as a selection effect of blue, luminous quasars and not as a general physical property of the model. This implies that SDSS quasars, with their narrow distribution of Eddington ratios, are not an ideal sample to study the dependencies of the observable spectral properties on the Eddington ratio. Such dependencies may be very strong and yet not be detectable in this sample.

\section{Summary and conclusions}
In this paper we started from the evidence that the Eddington ratio \ledd distribution for SDSS quasars based on virial black hole masses has a width of the same order of, if not slightly larger than, the expected uncertainty on black hole masses estimates. This suggests that the intrinsic distribution of \ledd may be very narrow. We then assumed, as a working hypothesis, that \ledd has a constant value of 0.1 for all the quasars in the SDSS DR16 catalogue, and that the accretion disk is well described by a standard optically thick, geometrically thin disk model. We explored the consequences of such assumptions and we obtained the following results:\\
1) We estimated the disk spectra assuming \ledd=0.1 and the observed luminosity. We found that nearly all the quasars in the SDSS DR16 sample have spectra that peak beyond the Lyman break, which are very similar to each other in the observable range. We stacked the few objects cold enough to have the peak in the observable range, and we indeed found the expected "cold" spectral shape (Figure~\ref{fig:moneyplot}). Instead, objects that should be "cold enough" according to their virial mass do not show any flattening at high frequencies.\\
2) We used the He~II~$\lambda 1640$~\AA line as a proxy of the ionizing continuum and we studied the dependence of its equivalent width on the FWHM of the Mg~II line and the continuum luminosity (Figure~\ref{fig:L_W_He2}). A PCA analysis revealed that the relation is correctly reproduced in the constant-\ledd scenario, while alternative assumptions, such as a standard model with black hole virial masses, or the model of \citet{hopkins2025} are not in agreement with the observations.\\
3) We assumed that the main broad emission lines in the optical-UV spectrum of quasars are a good proxy of the ionizing luminosity and we studied the line-continuum luminosity relations (i.e. the `Baldwin effect'). We found that the slopes of these relations are those expected in a standard disk scenario. Furthermore, we obtained a very small dispersion for each of the tested line--continuum relations (down to $\sim$0.13~dex). In general, this suggests that the line--continuum relation is driven by just one physical parameter. Specifically, within the standard disk model, it implies that the luminosity is almost only dependent on the black hole mass, and therefore the Eddington ratio distribution must be very tight.\\
4) While our estimate of the black hole mass for an individual quasar can be quite different from the virial estimate, the distribution of black hole masses obtained with the two methods are not systematically different from each other (Fig.~\ref{fig:Ledd-MBH}). 

Even if our analysis shows that there is little room for a significant width of the distribution of \ledd, our working hypothesis of \ledd=0.1 is obviously an oversimplification. 
We can estimate an upper limit of its real width from the results we have discussed: -from the distribution of the "virial" \ledd (Figure~\ref{fig:Ledd-Lbol}), considering that the individual contributions are added in quadrature, we can only predict $\sigma$(\ledd)$\lesssim0.2$~dex; -from the tightness of the line-continuum relations, considering that from Equation~\ref{eq:tmax}, the contribution of \ledd is with a power of 0.5, and that other contributions must be present (see the discussion in Section~\ref{sec:baldwin}), we can predict $\sigma$(\ledd)$\lesssim0.1-0.15$~dex.

If we assume log(\ledd)=$-1\pm0.15$ all our findings based on the simplified assumption \ledd=0.1 remain valid, including (a) a more precise estimate of the black hole mass by assuming a constant \ledd rather than using the viral estimate, and (b) the nearly complete absence of super-Eddington quasars in the SDSS-DR16 catalogue ($\sim 10^{-9}~\%$ in a Gaussian approximation). Regarding the latter point, we stress again that this does not imply that super-Eddington quasars do not exist, but only that they are not present in a sample of optically-selected, blue type~1 quasars such as the ones in our sample. The same caveat applies to the assumed accretion model: different choices may well provide a better representation of the quasar emission at different (either smaller or higher) accretion rates.

We conclude by noting that our proposed scenario (a "standard" accretion disk with a nearly constant Eddington ratio) provides an explanation to several observational facts by reducing the complexity of the model (in fact, removing one degree of freedom). This suggests that the observed uniformity of SDSS quasars may be indeed due to an intrinsic similarity of their power engine.

\bibliographystyle{aa}
\bibliography{bib}

@article{shakura1973,
       author = {{Shakura}, N.~I. and {Sunyaev}, R.~A.},
        title = "{Black holes in binary systems. Observational appearance.}",
      journal = {\aap},
         year = 1973,
        month = jan,
       volume = {24},
        pages = {337-355},
       adsurl = {https://ui.adsabs.harvard.edu/abs/1973A&A....24..337S}
}

@INPROCEEDINGS{novikov1973,
       author = {{Novikov}, I.~D. and {Thorne}, K.~S.},
        title = "{Astrophysics of black holes.}",
    booktitle = {Black Holes (Les Astres Occlus)},
         year = 1973,
       editor = {{Dewitt}, C. and {Dewitt}, B.~S.},
        month = jan,
        pages = {343-450},
       adsurl = {https://ui.adsabs.harvard.edu/abs/1973blho.conf..343N}
}

@article{abramowicz1988,
       author = {{Abramowicz}, M.~A. and {Czerny}, B. and {Lasota}, J.~P. and {Szuszkiewicz}, E.},
        title = "{Slim Accretion Disks}",
      journal = {\apj},
         year = 1988,
        month = sep,
       volume = {332},
        pages = {646},
          doi = {10.1086/166683},
       adsurl = {https://ui.adsabs.harvard.edu/abs/1988ApJ...332..646A}
}

@article{mitchell2023,
       author = {{Mitchell}, Jake A.~J. and {Done}, Chris and {Ward}, Martin J. and {Kynoch}, Daniel and {Hagen}, Scott and {Lusso}, Elisabeta and {Landt}, Hermine},
        title = "{The SOUX AGN sample: optical/UV/X-ray SEDs and the nature of the disc}",
      journal = {\mnras},
         year = 2023,
        month = sep,
       volume = {524},
       number = {2},
        pages = {1796-1825},
          doi = {10.1093/mnras/stad1830},
archivePrefix = {arXiv},
       eprint = {2210.11977},
 primaryClass = {astro-ph.GA},
       adsurl = {https://ui.adsabs.harvard.edu/abs/2023MNRAS.524.1796M}
}

@article{cai2023,
       author = {{Cai}, Zhen-Yi and {Wang}, Jun-Xian},
        title = "{A universal average spectral energy distribution for quasars from the optical to the extreme ultraviolet}",
      journal = {Nature Astronomy},
         year = 2023,
        month = dec,
       volume = {7},
        pages = {1506-1516},
          doi = {10.1038/s41550-023-02088-5},
archivePrefix = {arXiv},
       eprint = {2309.01541},
 primaryClass = {astro-ph.GA},
       adsurl = {https://ui.adsabs.harvard.edu/abs/2023NatAs...7.1506C}
}

@article{trefoloni2024,
       author = {{Trefoloni}, Bartolomeo and {Lusso}, Elisabeta and {Nardini}, Emanuele and {Risaliti}, Guido and {Marconi}, Alessandro and {Bargiacchi}, Giada and {Sacchi}, Andrea and {Pietrini}, Paola and {Signorini}, Matilde},
        title = "{Quasars as standard candles: VI. Spectroscopic validation of the cosmological sample}",
      journal = {\aap},
         year = 2024,
        month = sep,
       volume = {689},
          eid = {A109},
        pages = {A109},
          doi = {10.1051/0004-6361/202348938},
archivePrefix = {arXiv},
       eprint = {2404.07205},
 primaryClass = {astro-ph.GA},
       adsurl = {https://ui.adsabs.harvard.edu/abs/2024A&A...689A.109T}
}

@article{wu2022,
       author = {{Wu}, Qiaoya and {Shen}, Yue},
        title = "{A Catalog of Quasar Properties from Sloan Digital Sky Survey Data Release 16}",
      journal = {\apjs},
         year = 2022,
        month = dec,
       volume = {263},
       number = {2},
          eid = {42},
        pages = {42},
          doi = {10.3847/1538-4365/ac9ead},
archivePrefix = {arXiv},
       eprint = {2209.03987},
 primaryClass = {astro-ph.GA},
       adsurl = {https://ui.adsabs.harvard.edu/abs/2022ApJS..263...42W}
}

@article{baldwin1977,
       author = {{Baldwin}, Jack A.},
        title = "{Luminosity Indicators in the Spectra of Quasi-Stellar Objects}",
      journal = {\apj},
         year = 1977,
        month = jun,
       volume = {214},
        pages = {679-684},
          doi = {10.1086/155294},
       adsurl = {https://ui.adsabs.harvard.edu/abs/1977ApJ...214..679B}
}

@article{soltan1982,
       author = {{Soltan}, A.},
        title = "{Masses of quasars.}",
      journal = {\mnras},
         year = 1982,
        month = jul,
       volume = {200},
        pages = {115-122},
          doi = {10.1093/mnras/200.1.115},
       adsurl = {https://ui.adsabs.harvard.edu/abs/1982MNRAS.200..115S}
}

@article{marconi2004,
       author = {{Marconi}, A. and {Risaliti}, G. and {Gilli}, R. and {Hunt}, L.~K. and {Maiolino}, R. and {Salvati}, M.},
        title = "{Local supermassive black holes, relics of active galactic nuclei and the X-ray background}",
      journal = {\mnras},
         year = 2004,
        month = jun,
       volume = {351},
       number = {1},
        pages = {169-185},
          doi = {10.1111/j.1365-2966.2004.07765.x},
archivePrefix = {arXiv},
       eprint = {astro-ph/0311619},
 primaryClass = {astro-ph},
       adsurl = {https://ui.adsabs.harvard.edu/abs/2004MNRAS.351..169M}
}

@article{laor2014,
       author = {{Laor}, Ari and {Davis}, Shane W.},
        title = "{Line-driven winds and the UV turnover in AGN accretion discs}",
      journal = {\mnras},
         year = 2014,
        month = mar,
       volume = {438},
       number = {4},
        pages = {3024-3038},
          doi = {10.1093/mnras/stt2408},
archivePrefix = {arXiv},
       eprint = {1312.3556},
 primaryClass = {astro-ph.HE},
       adsurl = {https://ui.adsabs.harvard.edu/abs/2014MNRAS.438.3024L}
}

@article{lawrence2018,
       author = {{Lawrence}, Andy},
        title = "{Quasar viscosity crisis}",
      journal = {Nature Astronomy},
         year = 2018,
        month = feb,
       volume = {2},
        pages = {102-103},
          doi = {10.1038/s41550-017-0372-1},
archivePrefix = {arXiv},
       eprint = {1802.00408},
 primaryClass = {astro-ph.HE},
       adsurl = {https://ui.adsabs.harvard.edu/abs/2018NatAs...2..102L}
}

@article{morgan2010,
       author = {{Morgan}, Christopher W. and {Kochanek}, C.~S. and {Morgan}, Nicholas D. and {Falco}, Emilio E.},
        title = "{The Quasar Accretion Disk Size-Black Hole Mass Relation}",
      journal = {\apj},
         year = 2010,
        month = apr,
       volume = {712},
       number = {2},
        pages = {1129-1136},
          doi = {10.1088/0004-637X/712/2/1129},
archivePrefix = {arXiv},
       eprint = {1002.4160},
 primaryClass = {astro-ph.CO},
       adsurl = {https://ui.adsabs.harvard.edu/abs/2010ApJ...712.1129M}
}

@article{fausnaugh2016,
       author = {{Fausnaugh}, M.~M. and {Denney}, K.~D. and {Barth}, A.~J. and {Bentz}, M.~C. and {Bottorff}, M.~C. and {Carini}, M.~T. and {Croxall}, K.~V. and {De Rosa}, G. and {Goad}, M.~R. and {Horne}, Keith and {Joner}, M.~D. and {Kaspi}, S. and {Kim}, M. and {Klimanov}, S.~A. and {Kochanek}, C.~S. and {Leonard}, D.~C. and {Netzer}, H. and {Peterson}, B.~M. and {Schn{\"u}lle}, K. and {Sergeev}, S.~G. and {Vestergaard}, M. and {Zheng}, W.-K. and {Zu}, Y. and {Anderson}, M.~D. and {Ar{\'e}valo}, P. and {Bazhaw}, C. and {Borman}, G.~A. and {Boroson}, T.~A. and {Brandt}, W.~N. and {Breeveld}, A.~A. and {Brewer}, B.~J. and {Cackett}, E.~M. and {Crenshaw}, D.~M. and {Dalla Bont{\`a}}, E. and {De Lorenzo-C{\'a}ceres}, A. and {Dietrich}, M. and {Edelson}, R. and {Efimova}, N.~V. and {Ely}, J. and {Evans}, P.~A. and {Filippenko}, A.~V. and {Flatland}, K. and {Gehrels}, N. and {Geier}, S. and {Gelbord}, J.~M. and {Gonzalez}, L. and {Gorjian}, V. and {Grier}, C.~J. and {Grupe}, D. and {Hall}, P.~B. and {Hicks}, S. and {Horenstein}, D. and {Hutchison}, T. and {Im}, M. and {Jensen}, J.~J. and {Jones}, J. and {Kaastra}, J. and {Kelly}, B.~C. and {Kennea}, J.~A. and {Kim}, S.~C. and {Korista}, K.~T. and {Kriss}, G.~A. and {Lee}, J.~C. and {Lira}, P. and {MacInnis}, F. and {Manne-Nicholas}, E.~R. and {Mathur}, S. and {McHardy}, I.~M. and {Montouri}, C. and {Musso}, R. and {Nazarov}, S.~V. and {Norris}, R.~P. and {Nousek}, J.~A. and {Okhmat}, D.~N. and {Pancoast}, A. and {Papadakis}, I. and {Parks}, J.~R. and {Pei}, L. and {Pogge}, R.~W. and {Pott}, J.-U. and {Rafter}, S.~E. and {Rix}, H.-W. and {Saylor}, D.~A. and {Schimoia}, J.~S. and {Siegel}, M. and {Spencer}, M. and {Starkey}, D. and {Sung}, H.-I. and {Teems}, K.~G. and {Treu}, T. and {Turner}, C.~S. and {Uttley}, P. and {Villforth}, C. and {Weiss}, Y. and {Woo}, J.-H. and {Yan}, H. and {Young}, S.},
        title = "{Space Telescope and Optical Reverberation Mapping Project. III. Optical Continuum Emission and Broadband Time Delays in NGC 5548}",
      journal = {\apj},
         year = 2016,
        month = apr,
       volume = {821},
       number = {1},
          eid = {56},
        pages = {56},
          doi = {10.3847/0004-637X/821/1/56},
archivePrefix = {arXiv},
       eprint = {1510.05648},
 primaryClass = {astro-ph.GA},
       adsurl = {https://ui.adsabs.harvard.edu/abs/2016ApJ...821...56F}
}

@article{stockman1979,
       author = {{Stockman}, H.~S. and {Angel}, J.~R.~P. and {Miley}, G.~K.},
        title = "{Alignment of the optical polarization with the radio structure of QSOs.}",
      journal = {\apjl},
         year = 1979,
        month = jan,
       volume = {227},
        pages = {L55-L58},
          doi = {10.1086/182866},
       adsurl = {https://ui.adsabs.harvard.edu/abs/1979ApJ...227L..55S}
}

@INPROCEEDINGS{antonucci1988,
       author = {{Antonucci}, Robert},
        title = "{Polarization of Active Galactic Nuclei and Quasars}",
    booktitle = {Supermassive Black Holes},
         year = 1988,
       editor = {{Kafatos}, Minas},
        month = jan,
        pages = {26},
       adsurl = {https://ui.adsabs.harvard.edu/abs/1988smbh.proc...26A}
}

@article{york2000,
       author = {{York}, Donald G. and {Adelman}, J. and {Anderson}, Jr., John E. and {Anderson}, Scott F. and {Annis}, James and {Bahcall}, Neta A. and {Bakken}, J.~A. and {Barkhouser}, Robert and {Bastian}, Steven and {Berman}, Eileen and {Boroski}, William N. and {Bracker}, Steve and {Briegel}, Charlie and {Briggs}, John W. and {Brinkmann}, J. and {Brunner}, Robert and {Burles}, Scott and {Carey}, Larry and {Carr}, Michael A. and {Castander}, Francisco J. and {Chen}, Bing and {Colestock}, Patrick L. and {Connolly}, A.~J. and {Crocker}, J.~H. and {Csabai}, Istv{\'a}n and {Czarapata}, Paul C. and {Davis}, John Eric and {Doi}, Mamoru and {Dombeck}, Tom and {Eisenstein}, Daniel and {Ellman}, Nancy and {Elms}, Brian R. and {Evans}, Michael L. and {Fan}, Xiaohui and {Federwitz}, Glenn R. and {Fiscelli}, Larry and {Friedman}, Scott and {Frieman}, Joshua A. and {Fukugita}, Masataka and {Gillespie}, Bruce and {Gunn}, James E. and {Gurbani}, Vijay K. and {de Haas}, Ernst and {Haldeman}, Merle and {Harris}, Frederick H. and {Hayes}, J. and {Heckman}, Timothy M. and {Hennessy}, G.~S. and {Hindsley}, Robert B. and {Holm}, Scott and {Holmgren}, Donald J. and {Huang}, Chi-hao and {Hull}, Charles and {Husby}, Don and {Ichikawa}, Shin-Ichi and {Ichikawa}, Takashi and {Ivezi{\'c}}, {\v{Z}}eljko and {Kent}, Stephen and {Kim}, Rita S.~J. and {Kinney}, E. and {Klaene}, Mark and {Kleinman}, A.~N. and {Kleinman}, S. and {Knapp}, G.~R. and {Korienek}, John and {Kron}, Richard G. and {Kunszt}, Peter Z. and {Lamb}, D.~Q. and {Lee}, B. and {Leger}, R. French and {Limmongkol}, Siriluk and {Lindenmeyer}, Carl and {Long}, Daniel C. and {Loomis}, Craig and {Loveday}, Jon and {Lucinio}, Rich and {Lupton}, Robert H. and {MacKinnon}, Bryan and {Mannery}, Edward J. and {Mantsch}, P.~M. and {Margon}, Bruce and {McGehee}, Peregrine and {McKay}, Timothy A. and {Meiksin}, Avery and {Merelli}, Aronne and {Monet}, David G. and {Munn}, Jeffrey A. and {Narayanan}, Vijay K. and {Nash}, Thomas and {Neilsen}, Eric and {Neswold}, Rich and {Newberg}, Heidi Jo and {Nichol}, R.~C. and {Nicinski}, Tom and {Nonino}, Mario and {Okada}, Norio and {Okamura}, Sadanori and {Ostriker}, Jeremiah P. and {Owen}, Russell and {Pauls}, A. George and {Peoples}, John and {Peterson}, R.~L. and {Petravick}, Donald and {Pier}, Jeffrey R. and {Pope}, Adrian and {Pordes}, Ruth and {Prosapio}, Angela and {Rechenmacher}, Ron and {Quinn}, Thomas R. and {Richards}, Gordon T. and {Richmond}, Michael W. and {Rivetta}, Claudio H. and {Rockosi}, Constance M. and {Ruthmansdorfer}, Kurt and {Sandford}, Dale and {Schlegel}, David J. and {Schneider}, Donald P. and {Sekiguchi}, Maki and {Sergey}, Gary and {Shimasaku}, Kazuhiro and {Siegmund}, Walter A. and {Smee}, Stephen and {Smith}, J. Allyn and {Snedden}, S. and {Stone}, R. and {Stoughton}, Chris and {Strauss}, Michael A. and {Stubbs}, Christopher and {SubbaRao}, Mark and {Szalay}, Alexander S. and {Szapudi}, Istvan and {Szokoly}, Gyula P. and {Thakar}, Anirudda R. and {Tremonti}, Christy and {Tucker}, Douglas L. and {Uomoto}, Alan and {Vanden Berk}, Dan and {Vogeley}, Michael S. and {Waddell}, Patrick and {Wang}, Shu-i. and {Watanabe}, Masaru and {Weinberg}, David H. and {Yanny}, Brian and {Yasuda}, Naoki and {SDSS Collaboration}},
        title = "{The Sloan Digital Sky Survey: Technical Summary}",
      journal = {\aj},
         year = 2000,
        month = sep,
       volume = {120},
       number = {3},
        pages = {1579-1587},
          doi = {10.1086/301513},
archivePrefix = {arXiv},
       eprint = {astro-ph/0006396},
 primaryClass = {astro-ph},
       adsurl = {https://ui.adsabs.harvard.edu/abs/2000AJ....120.1579Y}
}

@article{kubota2018,
       author = {{Kubota}, Aya and {Done}, Chris},
        title = "{A physical model of the broad-band continuum of AGN and its implications for the UV/X relation and optical variability}",
      journal = {\mnras},
         year = 2018,
        month = oct,
       volume = {480},
       number = {1},
        pages = {1247-1262},
          doi = {10.1093/mnras/sty1890},
archivePrefix = {arXiv},
       eprint = {1804.00171},
 primaryClass = {astro-ph.HE},
       adsurl = {https://ui.adsabs.harvard.edu/abs/2018MNRAS.480.1247K}
}

@article{vestergaard2006,
       author = {{Vestergaard}, Marianne and {Peterson}, Bradley M.},
        title = "{Determining Central Black Hole Masses in Distant Active Galaxies and Quasars. II. Improved Optical and UV Scaling Relationships}",
      journal = {\apj},
         year = 2006,
        month = apr,
       volume = {641},
       number = {2},
        pages = {689-709},
          doi = {10.1086/500572},
archivePrefix = {arXiv},
       eprint = {astro-ph/0601303},
 primaryClass = {astro-ph},
       adsurl = {https://ui.adsabs.harvard.edu/abs/2006ApJ...641..689V}
}

@article{park2012,
       author = {{Park}, Daeseong and {Woo}, Jong-Hak and {Treu}, Tommaso and {Barth}, Aaron J. and {Bentz}, Misty C. and {Bennert}, Vardha N. and {Canalizo}, Gabriela and {Filippenko}, Alexei V. and {Gates}, Elinor and {Greene}, Jenny E. and {Malkan}, Matthew A. and {Walsh}, Jonelle},
        title = "{The Lick AGN Monitoring Project: Recalibrating Single-epoch Virial Black Hole Mass Estimates}",
      journal = {\apj},
         year = 2012,
        month = mar,
       volume = {747},
       number = {1},
          eid = {30},
        pages = {30},
          doi = {10.1088/0004-637X/747/1/30},
archivePrefix = {arXiv},
       eprint = {1111.6604},
 primaryClass = {astro-ph.CO},
       adsurl = {https://ui.adsabs.harvard.edu/abs/2012ApJ...747...30P}
}

@article{dallabonta2020,
       author = {{Dalla Bont{\`a}}, Elena and {Peterson}, Bradley M. and {Bentz}, Misty C. and {Brandt}, W.~N. and {Ciroi}, S. and {De Rosa}, Gisella and {Fonseca Alvarez}, Gloria and {Grier}, Catherine J. and {Hall}, P.~B. and {Hern{\'a}ndez Santisteban}, Juan V. and {Ho}, Luis C. and {Homayouni}, Y. and {Horne}, Keith and {Kochanek}, C.~S. and {Li}, Jennifer I.-Hsiu and {Morelli}, L. and {Pizzella}, A. and {Pogge}, R.~W. and {Schneider}, D.~P. and {Shen}, Yue and {Trump}, J.~R. and {Vestergaard}, Marianne},
        title = "{The Sloan Digital Sky Survey Reverberation Mapping Project: Estimating Masses of Black Holes in Quasars with Single-epoch Spectroscopy}",
      journal = {\apj},
         year = 2020,
        month = nov,
       volume = {903},
       number = {2},
          eid = {112},
        pages = {112},
          doi = {10.3847/1538-4357/abbc1c},
archivePrefix = {arXiv},
       eprint = {2007.02963},
 primaryClass = {astro-ph.GA},
       adsurl = {https://ui.adsabs.harvard.edu/abs/2020ApJ...903..112D}
}

@article{shen2013,
      author = {{Shen}, Yue},
        title = "{The mass of quasars}",
      journal = {Bulletin of the Astronomical Society of India},
         year = 2013,
        month = mar,
       volume = {41},
       number = {1},
        pages = {61-115},
          doi = {10.48550/arXiv.1302.2643},
archivePrefix = {arXiv},
       eprint = {1302.2643},
 primaryClass = {astro-ph.CO},
       adsurl = {https://ui.adsabs.harvard.edu/abs/2013BASI...41...61S}
}

@article{signorini2024,
       author = {{Signorini}, Matilde and {Risaliti}, Guido and {Lusso}, Elisabeta and {Nardini}, Emanuele and {Bargiacchi}, Giada and {Sacchi}, Andrea and {Trefoloni}, Bartolomeo},
        title = "{Quasars as standard candles. V. Accounting for the dispersion in the L$_{X}$-L$_{UV}$ relation down to {\ensuremath{\leq}} 0.06 dex}",
      journal = {\aap},
         year = 2024,
        month = jul,
       volume = {687},
          eid = {A32},
        pages = {A32},
          doi = {10.1051/0004-6361/202348941},
archivePrefix = {arXiv},
       eprint = {2312.08448},
 primaryClass = {astro-ph.CO},
       adsurl = {https://ui.adsabs.harvard.edu/abs/2024A&A...687A..32S}
}

@article{shen2024,
       author = {{Shen}, Yue and {Grier}, Catherine J. and {Horne}, Keith and {Stone}, Zachary and {Li}, Jennifer I. and {Yang}, Qian and {Homayouni}, Yasaman and {Trump}, Jonathan R. and {Anderson}, Scott F. and {Brandt}, W.~N. and {Hall}, Patrick B. and {Ho}, Luis C. and {Jiang}, Linhua and {Petitjean}, Patrick and {Schneider}, Donald P. and {Tao}, Charling and {Donnan}, Fergus. R. and {AlSayyad}, Yusra and {Bershady}, Matthew A. and {Blanton}, Michael R. and {Bizyaev}, Dmitry and {Bundy}, Kevin and {Chen}, Yuguang and {Davis}, Megan C. and {Dawson}, Kyle and {Fan}, Xiaohui and {Greene}, Jenny E. and {Gr{\"o}ller}, Hannes and {Guo}, Yucheng and {Ibarra-Medel}, H{\'e}ctor and {Jiang}, Yuanzhe and {Keenan}, Ryan P. and {Kollmeier}, Juna A. and {Lejoly}, Cassandra and {Li}, Zefeng and {de la Macorra}, Axel and {Moe}, Maxwell and {Nie}, Jundan and {Rossi}, Graziano and {Smith}, Paul S. and {Tee}, Wei Leong and {Weijmans}, Anne-Marie and {Xu}, Jiachuan and {Yue}, Minghao and {Zhou}, Xu and {Zhou}, Zhimin and {Zou}, Hu},
        title = "{The Sloan Digital Sky Survey Reverberation Mapping Project: Key Results}",
      journal = {\apjs},
         year = 2024,
        month = jun,
       volume = {272},
       number = {2},
          eid = {26},
        pages = {26},
          doi = {10.3847/1538-4365/ad3936},
archivePrefix = {arXiv},
       eprint = {2305.01014},
 primaryClass = {astro-ph.GA},
       adsurl = {https://ui.adsabs.harvard.edu/abs/2024ApJS..272...26S}
}

@article{hagen2023,
       author = {{Hagen}, Scott and {Done}, Chris},
        title = "{Estimating black hole spin from AGN SED fitting: the impact of general-relativistic ray tracing}",
      journal = {\mnras},
         year = 2023,
        month = nov,
       volume = {525},
       number = {3},
        pages = {3455-3467},
          doi = {10.1093/mnras/stad2499},
archivePrefix = {arXiv},
       eprint = {2304.01253},
 primaryClass = {astro-ph.HE},
       adsurl = {https://ui.adsabs.harvard.edu/abs/2023MNRAS.525.3455H}
}

@article{schlafly2011,
       author = {{Schlafly}, Edward F. and {Finkbeiner}, Douglas P.},
        title = "{Measuring Reddening with Sloan Digital Sky Survey Stellar Spectra and Recalibrating SFD}",
      journal = {\apj},
         year = 2011,
        month = aug,
       volume = {737},
       number = {2},
          eid = {103},
        pages = {103},
          doi = {10.1088/0004-637X/737/2/103},
archivePrefix = {arXiv},
       eprint = {1012.4804},
 primaryClass = {astro-ph.GA},
       adsurl = {https://ui.adsabs.harvard.edu/abs/2011ApJ...737..103S}
}

@article{fitzpatrick1999,
       author = {{Fitzpatrick}, Edward L.},
        title = "{Correcting for the Effects of Interstellar Extinction}",
      journal = {\pasp},
         year = 1999,
        month = jan,
       volume = {111},
       number = {755},
        pages = {63-75},
          doi = {10.1086/316293},
archivePrefix = {arXiv},
       eprint = {astro-ph/9809387},
 primaryClass = {astro-ph},
       adsurl = {https://ui.adsabs.harvard.edu/abs/1999PASP..111...63F}
}

@ARTICLE{hopkins2025,
       author = {{Hopkins}, Philip F.},
        title = "{Multi-Phase Thermal Structure \& The Origin of the Broad-Line Region, Torus, and Corona in Magnetically-Dominated Accretion Disks}",
      journal = {The Open Journal of Astrophysics},
         year = 2025,
        month = may,
       volume = {8},
          eid = {56},
        pages = {56},
          doi = {10.33232/001c.137969},
archivePrefix = {arXiv},
       eprint = {2407.00160},
 primaryClass = {astro-ph.GA},
       adsurl = {https://ui.adsabs.harvard.edu/abs/2025OJAp....8E..56H}
}

@article{lusso2015first,
  title={The first ultraviolet quasar-stacked spectrum at z≃ 2.4 from WFC3},
  author={Lusso, E and Worseck, G and Hennawi, JF and Prochaska, JX and Vignali, Cristian and Stern, J and O'Meara, JM},
  journal={MNRAS},
  volume={449},
  number={4},
  pages={4204--4220},
  year={2015},
  publisher={The Royal Astronomical Society}
}

@article{wolf2024accretion,
  title={The accretion of a solar mass per day by a 17-billion solar mass black hole},
  author={Wolf, Christian and Lai, Samuel and Onken, Christopher A and Amrutha, Neelesh and Bian, Fuyan and Hon, Wei Jeat and Tisserand, Patrick and Webster, Rachel L},
  journal={Nat Astronomy},
  volume={8},
  number={4},
  pages={520--529},
  year={2024},
  publisher={Nature Publishing Group UK London}
}

@article{ferland2020state,
  title={State-of-the-art AGN SEDs for photoionization models: BLR predictions confront the observations},
  author={Ferland, GJ and Done, Chris and Jin, Chichuan and Landt, Hermine and Ward, MJ},
  journal={MNRAS},
  volume={494},
  number={4},
  pages={5917--5922},
  year={2020},
  publisher={Oxford University Press}
}

@article{temple2023testing,
  title={Testing AGN outflow and accretion models with C iv and He ii emission line demographics in z≈ 2 quasars},
  author={Temple, Matthew J and Matthews, James H and Hewett, Paul C and Rankine, Amy L and Richards, Gordon T and Banerji, Manda and Ferland, Gary J and Knigge, Christian and Stepney, Matthew},
  journal={MNRAS},
  volume={523},
  number={1},
  pages={646--666},
  year={2023},
  publisher={Oxford University Press}
}

@article{trefoloni2025missing,
  title={The missing Fe II bump in faint JWST active galactic nuclei: Possible evidence of metal-poor broad-line regions at early cosmic times},
  author={Trefoloni, Bartolomeo and Ji, Xihan and Maiolino, Roberto and D’Eugenio, Francesco and {\"U}bler, Hannah and Scholtz, Jan and Marconi, Alessandro and Marconcini, Cosimo and Mazzolari, Giovanni},
  journal={A\&A},
  volume={700},
  pages={A203},
  year={2025},
  publisher={EDP Sciences}
}

@article{richards2006spectral,
  title={Spectral energy distributions and multiwavelength selection of type 1 quasars},
  author={Richards, Gordon T and Lacy, Mark and Storrie-Lombardi, Lisa J and Hall, Patrick B and Gallagher, SC and Hines, Dean C and Fan, Xiaohui and Papovich, Casey and Vanden Berk, Daniel E and Trammell, George B and others},
  journal={ApJS},
  volume={166},
  number={2},
  pages={470},
  year={2006},
  publisher={IOP Publishing}
}

@article{risaliti2019cosmological,
  title={Cosmological constraints from the Hubble diagram of quasars at high redshifts},
  author={Risaliti, Guido and Lusso, Elisabeta},
  journal={Nat Astronomy},
  volume={3},
  number={3},
  pages={272--277},
  year={2019},
  publisher={Nature Publishing Group UK London}
}

@article{foreman2013emcee,
  title={emcee: the MCMC hammer},
  author={Foreman-Mackey, Daniel and Hogg, David W and Lang, Dustin and Goodman, Jonathan},
  journal={PASP},
  volume={125},
  number={925},
  pages={306},
  year={2013},
  publisher={IOP Publishing}
}

@article{trefoloni2025accretion,
  title={The accretion of quasars at the epoch of reionisation: $ JWST $ catches the primeval monsters slowly feasting},
  author={Trefoloni, Bartolomeo and Nardini, Emanuele and Carniani, Stefano and Lusso, Elisabeta and Marconi, Alessandro and Parlanti, Eleonora and Sacchi, Andrea and Shlentsova, Anastasia and Signorini, Matilde and Risaliti, Guido and others},
  journal={arXiv preprint arXiv:2512.16981},
  year={2025}
}

@article{netzer2019bolometric,
  title={Bolometric correction factors for active galactic nuclei},
  author={Netzer, Hagai},
  journal={MNRAS},
  volume={488},
  number={4},
  pages={5185--5191},
  year={2019},
  publisher={Oxford University Press}
}

@article{shen2011catalog,
  title={A catalog of quasar properties from sloan digital sky survey data release 7},
  author={Shen, Yue and Richards, Gordon T and Strauss, Michael A and Hall, Patrick B and Schneider, Donald P and Snedden, Stephanie and Bizyaev, Dmitry and Brewington, Howard and Malanushenko, Viktor and Malanushenko, Elena and others},
  journal={ApJS},
  volume={194},
  number={2},
  pages={45},
  year={2011},
  publisher={IOP Publishing}
}

@article{maiolino2025jwst,
  title={JWST meets Chandra: a large population of Compton thick, feedback-free, and intrinsically X-ray weak AGN, with a sprinkle of SNe},
  author={Maiolino, Roberto and Risaliti, Guido and Signorini, Matilde and Trefoloni, Bartolomeo and Juod{\v{z}}balis, Ignas and Scholtz, Jan and {\"U}bler, Hannah and D’Eugenio, Francesco and Carniani, Stefano and Fabian, Andy and others},
  journal={Monthly Notices of the Royal Astronomical Society},
  volume={538},
  number={3},
  pages={1921--1943},
  year={2025},
  publisher={Oxford University Press}
}

@article{matthee2024little,
  title={Little red dots: an abundant population of faint active galactic nuclei at z~ 5 revealed by the EIGER and FRESCO JWST surveys},
  author={Matthee, Jorryt and Naidu, Rohan P and Brammer, Gabriel and Chisholm, John and Eilers, Anna-Christina and Goulding, Andy and Greene, Jenny and Kashino, Daichi and Labbe, Ivo and Lilly, Simon J and others},
  journal={The Astrophysical Journal},
  volume={963},
  number={2},
  pages={129},
  year={2024},
  publisher={IOP Publishing}
}

@article{li2005multitemperature,
  title={Multitemperature blackbody spectrum of a thin accretion disk around a Kerr black hole: model computations and comparison with observations},
  author={Li, Li-Xin and Zimmerman, Erik R and Narayan, Ramesh and McClintock, Jeffrey E},
  journal={The Astrophysical Journal Supplement Series},
  volume={157},
  number={2},
  pages={335},
  year={2005},
  publisher={IOP Publishing}
}

\begin{appendix}
\section{Additional Redshift Effects}
\label{app:supp_figures}
Here we present an additional figure useful to complement the analysis.

\begin{figure}[h!]
\centering
\includegraphics[width=\linewidth,clip]{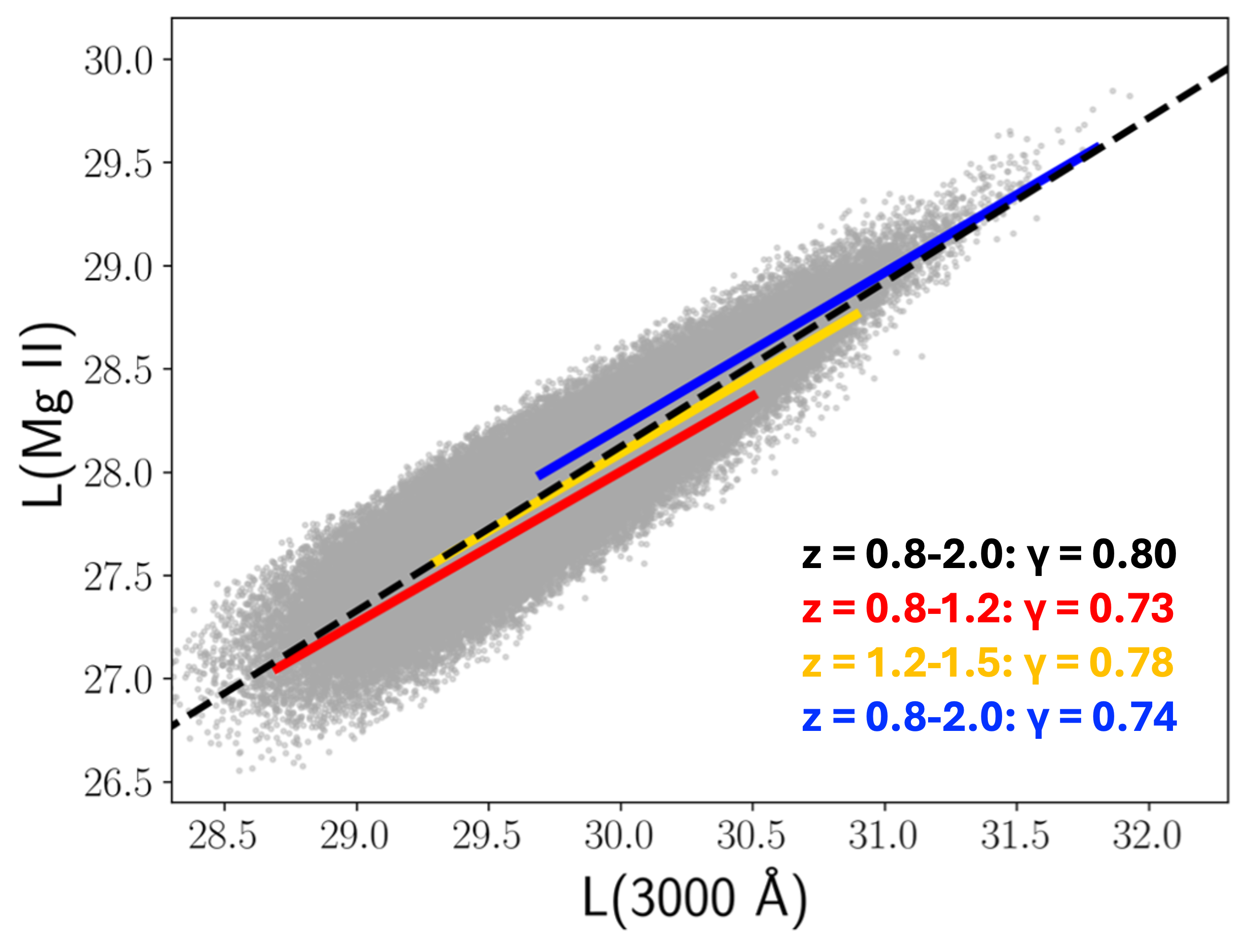}
\caption{Mg~II continuum--luminosity plot for the whole SDSS sample in the redshift range 0.8-2.0. The black dashed line shows the log-linear best fit, with a slope $\gamma=0.80$. The red, yellow and blue lines show the best fits for the three subsamples in the redshift intervals 0.8-1.2, 1.2-1.5 and 1.5-2.0, respectively, with slopes $\gamma\sim0.75$. The plot illustrates the importance of analyzing the line-continuum relation in small redshift bins to recover the intrinsic slope of the relation.}
\label{fig:BaldwinLum}
\end{figure}

In Fig.\ref{fig:BaldwinLum}, we show a plot of the line-luminosity relation for the Mg~II~$\lambda 2800$~\AA\, line for the whole sample and in three different redshift intervals. This illustrates the need for an analysis in small redshift intervals in order to obtain the correct slope of the relation. As shown in the figure, the difference between the slopes in small redhift intervals and in the whole sample is due to a change in the normalizations of the individual best-fit lines. This can be due to two effects: either the distance estimates, based on the standard \lcdm "concordance" model, are incorrect at high redshifts, or there is an additional physical dependence on redshift of the normalizations. The latter effect is consistent with an average \ledd slightly increasing with redshift, as in Figure~1: the dependence on \ledd$^{1/2}$ in Equation~2 implies an increase in normalization of the relation $L_{\rm BOL}$ - $L_\nu$. 


\end{appendix}
\end{document}